\documentclass{article}

\usepackage{microtype}
\usepackage{graphicx}
\usepackage{subcaption}
\usepackage{booktabs} 

\usepackage{hyperref}

\usepackage[accepted]{icml2025}

\usepackage{amsmath}
\usepackage{amssymb}
\usepackage{mathtools}
\usepackage{amsthm}

\usepackage[capitalize,noabbrev]{cleveref}

\theoremstyle{plain}

\theoremstyle{definition}

\theoremstyle{remark}

\usepackage[textsize=tiny]{todonotes}

\icmltitlerunning{DiffPR: Diffusion-Based Phase Reconstruction via Frequency-Decoupled Learning}

\usepackage{amsmath,amsfonts,bm}

\def\1{\bm{1}}

\def\eps{{\epsilon}}

\def\cN{{\mathcal{N}}}

\def\vf{{\bm{f}}}

\def\vx{{\bm{x}}}

\def\mI{{\bm{I}}}

\DeclareMathAlphabet{\mathsfit}{\encodingdefault}{\sfdefault}{m}{sl}
\SetMathAlphabet{\mathsfit}{bold}{\encodingdefault}{\sfdefault}{bx}{n}

\newcommand{\E}{\mathbb{E}}

\makeatletter
\def\thickhline{%
  \noalign{\ifnum0=`}\fi\hrule \@height \thickarrayrulewidth \futurelet
   \reserved@a\@xthickhline}
\def\@xthickhline{\ifx\reserved@a\thickhline
               \vskip\doublerulesep
               \vskip-\thickarrayrulewidth
             \fi
      \ifnum0=`{\fi}}
\makeatother

\newlength{\thickarrayrulewidth}
\usepackage{xcolor}
\definecolor{reference}{HTML}{F37171}
\definecolor{ode}{HTML}{FBE3D6}
\definecolor{sde}{HTML}{CAEEFB}
\definecolor{train}{HTML}{D9F2D0}

\usepackage{enumitem}
\usepackage{multirow} 
\usepackage{booktabs} 
\usepackage{siunitx}  
\usepackage{makecell}
\usepackage{amsmath, amsfonts} 
\usepackage{algorithm}
\usepackage{algpseudocode}

\begin{document}

\twocolumn[
\icmltitle{DiffPR: Diffusion-Based Phase Reconstruction via Frequency-Decoupled Learning}



\icmlsetsymbol{equal}{*}

\begin{icmlauthorlist}
\icmlauthor{Yi Zhang}{yyy}
\end{icmlauthorlist}

\icmlaffiliation{yyy}{Institude of Data Science, University of Hong Kong}

\icmlcorrespondingauthor{Yi Zhang}{yizhang101@connect.hku.hk}

\icmlkeywords{Machine Learning, ICML}

\vskip 0.3in
]



\printAffiliationsAndNotice{}  

\textbf{Abstract}\quad
Oversmoothing remains a persistent problem when applying deep learning to off-axis quantitative phase imaging (QPI): end-to-end U-Nets favour low-frequency content and under-represent fine, diagnostic detail. We trace this issue to \emph{spectral bias} and show that the bias is reinforced by the high-level skip connections that feed high-frequency features directly into the decoder. Removing those deepest skips—thus supervising the network only at a low resolution—significantly improves generalisation and fidelity. Building on this insight, we introduce \textbf{DiffPR}, a two-stage frequency-decoupled framework.
\textit{Stage 1}: an \emph{asymmetric} U-Net with cancelled high-frequency skips predicts a $1/4$-scale phase map from the interferogram, capturing reliable low-frequency structure while avoiding spectral bias.
\textit{Stage 2}: the upsampled prediction, lightly perturbed with Gaussian noise, is refined by an unconditional diffusion model that iteratively recovers the missing high-frequency residuals through reverse denoising. Experiments on four QPI datasets (\textit{B-Cell}, \textit{WBC}, \textit{HeLa}, \textit{3T3}) show that DiffPR outperforms strong U-Net baselines, boosting PSNR by up to \SI{1.1}{dB} and reducing MAE by $11\%$, while delivering markedly sharper membrane ridges and speckle patterns. The results demonstrate that cancelling high-level skips and delegating detail synthesis to a diffusion prior is an effective remedy for the spectral bias that limits conventional phase-retrieval networks.
\section{Introduction}

\begin{figure}[t]
  \centering
  \includegraphics[width=\linewidth]{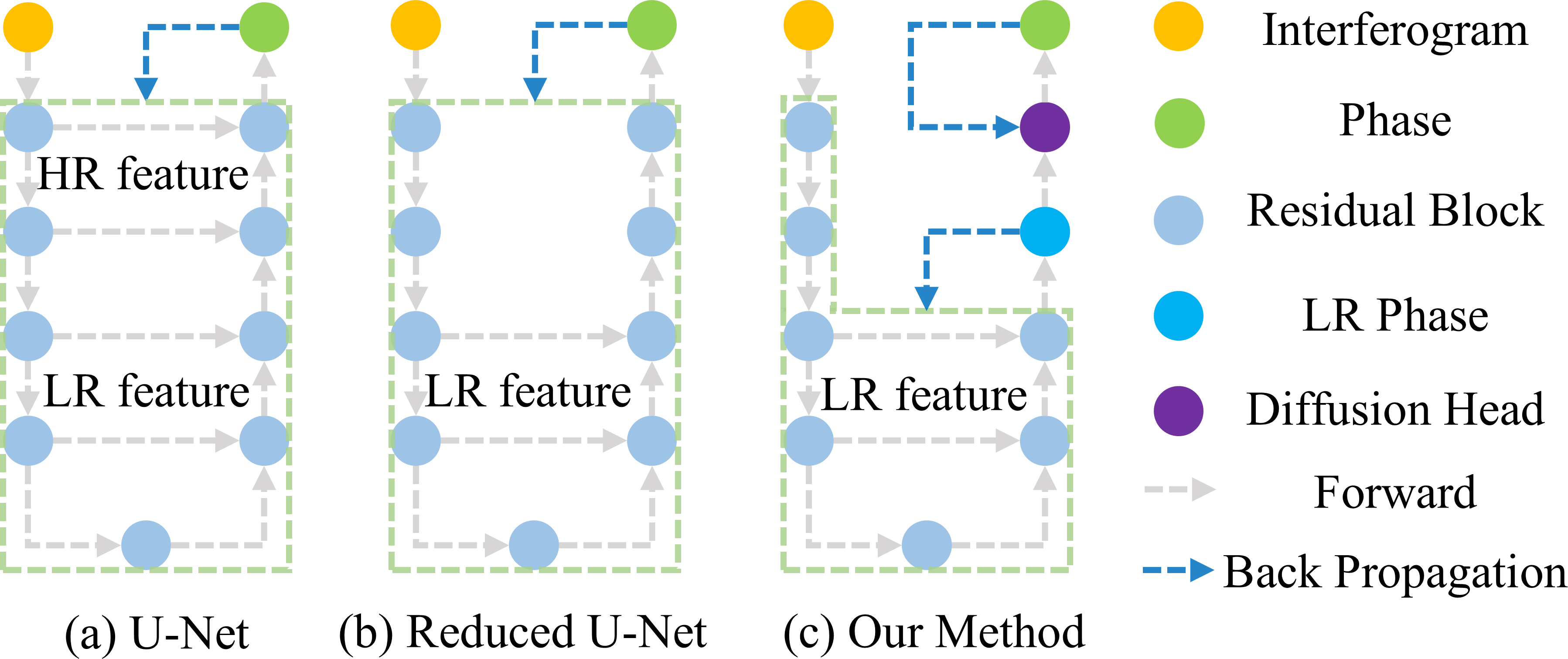} 
  \caption{\textbf{Network comparison.}
    (a)~Vanilla U\=/Net fuses both low- and high-resolution (HR) features via full skip connections;
    (b)~Reduced U\=/Net removes top-level skips, forcing the decoder to rely on low-resolution (LR) information only;
    (c)~Our DiffPR replaces the HR skip path with a diffusion head that synthesises high-frequency details from the predicted LR phase.
    Dashed blue arrows indicate gradients flowing from HR supervision.}
  \label{fig:method}
\end{figure}

\section{Introduction}
Off-axis quantitative phase imaging (QPI) provides high-sensitivity, label-free measurement of optical phase delay and has shown strong potential in applications such as live cell imaging \cite{Park:06,pandey2019integration}, morphology analysis \cite{ryu2021label,popescu2006diffraction}, and cellular mechanism studies \cite{pandey2019integration, popescu2006diffraction}. However, conventional phase reconstruction in off-axis QPI requires computationally expensive post-processing and background calibration, limiting its real-time deployment \cite{bhaduri2014diffraction,park2018quantitative,niu2020portable}. To address this, recent efforts have applied deep learning to directly map interferograms to phase maps \cite{b2,b3,b4,b5,b6}, enabling faster, calibration-free reconstruction.

Despite these advances, deep neural networks often struggle to recover fine structural details. A well-documented phenomenon in modern deep learning is the spectral bias: neural networks tend to learn low-frequency components significantly faster than high-frequency ones during training \cite{pmlr-v97-rahaman19a}. This mismatch becomes particularly problematic for high-resolution phase reconstruction, where high-frequency phase details are essential but difficult to learn directly.

In this work, we investigate how network design impacts frequency learning in deep learning-based phase reconstruction. Specifically, we revisit the common use of U-Net architectures \cite{ronneberger2015u, rivenson2018phase, zhang2021phasegan}, which fuse multi-level features through skip connections. While lower-level skip connections primarily carry low-frequency content, higher-level skip connections often reintroduce high-frequency information. Surprisingly, we find that removing these high-level skip connections (see Fig.~\ref{fig:method}~b), which suppresses direct high-frequency supervision, actually improves reconstruction fidelity and numerical performance. This suggests that simultaneously learning low- and high-frequency content, via architectural design, can be detrimental in this task.

Motivated by this observation, we propose DiffPR, a novel frequency-decoupled framework for phase reconstruction. DiffPR first uses a feedforward neural network to predict a low-resolution phase map from the interferogram. This low-resolution output is then upsampled to the target resolution and perturbed with mild Gaussian noise. Finally, an unconditional diffusion model transforms this noisy input into a high-resolution phase map through a reverse denoising process. Notably, diffusion models are inherently well-suited for learning high-frequency details. During training, they are explicitly tasked with recovering the original signal from inputs corrupted by small amounts of noise—effectively enforcing a learning bias toward fine-grained structures. This makes them particularly powerful for restoring the high-frequency components that standard networks struggle to learn. By decoupling frequency learning and leveraging this bias, our approach achieves sharper, more accurate reconstructions without the need for end-to-end high-frequency supervision.

Our contributions are summarized as follows:
\begin{enumerate}
    \item We empirically demonstrate that removing high-frequency skip connections in U-Net leads to improved phase reconstruction, revealing that direct joint learning of low- and high-frequency features can hinder performance.
    \item We introduce DiffPR, a diffusion-based framework that decouples frequency learning: a feedforward model reconstructs low-frequency phase content, and a diffusion model generates the high-frequency residuals, yielding high-fidelity phase maps with sharper structures.
\end{enumerate}

\section{Related Work}
\subsection{Phase Retrieval on Off-Axis Phase Retrieval
}
Off-axis quantitative phase imaging (QPI) has matured from its optical foundations into a versatile tool for label-free bio-imaging. Early diffraction-phase microscopy demonstrated sub-nanometer path-length sensitivity for live cells and laid the groundwork for subsequent biomedical applications \cite{Park:06,popescu2006diffraction,bhaduri2014diffraction}. Follow-up studies broadened QPI’s reach to morphology analysis, mechano-molecular phenotyping and portable systems, enabling in-field metrology and point-of-care diagnostics \cite{park2018quantitative,niu2020portable,pandey2019integration,ryu2021label}. Together, these efforts established the need for rapid, high-quality phase reconstruction pipelines to unlock QPI’s real-time potential.

To overcome the computational burden of Fourier-based or iterative solvers, the community has increasingly embraced deep learning. Rivenson \textit{et al.} first showed that a CNN could directly map interferograms to phase maps, suppressing twin-image artefacts with millisecond inference times \cite{rivenson2018phase}. Subsequent U-Net variants achieved real-time digital focusing and phase compensation in off-axis microscopy \cite{Zhang:18,Wang:18}, while one-step networks tackled the notoriously difficult phase-unwrapping task under heavy noise \cite{Wang:19,b2}. Extensions such as PhaseGAN introduced adversarial loss to mitigate the characteristic over-smoothing of $\ell_{1}$-trained models \cite{zhang2021phasegan}, whereas data-driven aberration modelling enabled truly calibration-free QPI \cite{b3}. Despite these advances, conventional encoder–decoder designs still struggle to recover high-frequency details, motivating our frequency-decoupled DiffPR framework, which predicts reliable low-frequency structure first and then synthesises high-frequency residuals with a generative prior.

\subsection{Diffusion Models and High-Frequency Detail Restoration}

Denoising Diffusion Probabilistic Models (DDPM) first framed generation as a gradual denoising process, showing that repeatedly predicting clean data from slightly noised inputs can synthesize sharp images~\cite{ho2020denoising}.  
Subsequent work demonstrated that diffusion models not only rival but surpass GANs in perceptual fidelity, especially on edge-rich datasets~\cite{dhariwal2021diffusion}.  
Nichol and Dhariwal further improved the noise schedule and sampling procedure, boosting detail preservation in generated outputs~\cite{nichol2021improved}.  Score-based generative modeling reformulated DDPMs as stochastic differential equations, providing theoretical insight into how successive denoising naturally reconstructs fine spatial frequencies~\cite{song2020score}.  
Latent Diffusion Models compress pixel space yet still reproduce high-frequency textures at megapixel resolution, underscoring the framework’s efficiency in retaining details after heavy dimensionality reduction~\cite{rombach2022latent}.  

Targeted restoration tasks highlight the bias toward high-frequency learning: SR3 recovers $\times8$ super-resolution facial details via iterative diffusion refinement~\cite{saharia2021sr3}, while \emph{Palette} generalises the paradigm to diverse image-to-image translations with texture-faithful outputs~\cite{saharia2022palette}.  
Variational Diffusion Models show that carefully chosen noise schedules and variational objectives yield state-of-the-art likelihoods without sacrificing edge sharpness~\cite{kingma2021vdm}, and EDM systematically analyses how sampler preconditioning accentuates high-frequency convergence~\cite{karras2022edm}.  

\subsection{Spectral Bias of Deep Neural Networks}

A growing body of work shows that deep networks possess a \emph{spectral bias}: during training they fit low-frequency components far earlier than high-frequency ones. Rahaman \emph{et al.}\ first quantified this effect with Fourier analysis, demonstrating that ReLU networks converge in a frequency-dependent manner and struggle to represent rapid oscillations without extensive parameter tuning~\cite{pmlr-v97-rahaman19a}. Xu \emph{et al.}\ verified the phenomenon on high-dimensional benchmarks and provided a theoretical explanation based on activation-function regularity~\cite{xu2019frequency}. To mitigate the bias, Tancik \emph{et al.}\ introduced Fourier feature positional encodings that pre-condition the neural tangent kernel and enable multilayer perceptrons to capture high-frequency signals~\cite{tancik2020fourier}.

Recent analyses interpret denoising diffusion probabilistic models (DDPMs) as \emph{approximate autoregressive generators in the frequency domain}: the reverse process first reconstructs low-frequency coefficients and then sequentially refines higher-frequency bins~\cite{dieleman2024spectral,gianluca2025spectral}. This spectral progression biases diffusion priors toward faithful high-frequency restoration, offering a natural complement to our DiffPR framework, which delegates high-frequency synthesis to an unconditional diffusion stage while maintaining reliable low-frequency structure through a feed-forward predictor.

\section{Method}




\begin{algorithm}[t]
  \caption{Inference pipeline of DiffPR}
  \label{alg:diffpr}
  \begin{algorithmic}[1]
    \Require interferogram $I$; predictor $f_{\theta}$; diffusion model $\epsilon_{\theta}$; noise schedule $\{\alpha_{t}\}_{t=1}^{T}$; corruption level $t_{\text{corrupt}}$; Reverse Diffusion Solver $\mathcal{R}$
    \State \textbf{Low-frequency prediction: }$\hat{\phi}^{\text{LR}}\leftarrow f_{\theta}(I)$
    \State \textbf{Initialisation:} 
          \(\tilde{\phi}_{T}\leftarrow \texttt{Up}\bigl(\hat{\phi}^{\text{LR}}\bigr)+
           \sigma_{t_{\text{corrupt}}}\,\epsilon,\quad \epsilon\sim\mathcal{N}(0,\mathbf{I})\)
    \For{$t=t_{\text{corrupt}},\dots,0$}
        \State \textbf{Predictor step:} 
              \(\tilde{\phi}_{t-1}\!\leftarrow\! \mathcal{R}(\tilde{\phi}_{t},t,\epsilon_\theta)\)
    \EndFor
    \State \Return \(\hat{\phi}^{\text{HR}}\leftarrow\tilde{\phi}_{0}\)
  \end{algorithmic}
\end{algorithm}

Our DiffPR framework consists of two stages:  
\textbf{(i)~Low-resolution phase prediction}, where an asymmetric U-Net maps the input interferogram to a \(\tfrac14\)-scale phase map;  
\textbf{(ii)~High-frequency refinement}, where a symmetric diffusion U-Net generates the final high-resolution phase by reversing a stochastic denoising process.  We denote an interferogram by \(I\in\mathbb{R}^{H\times W}\), the ground-truth phase by \(\phi^{\text{HR}}\), and the \(\tfrac14\)-scale phase by \(\phi^{\text{LR}}=\texttt{Down}(\phi^{\text{HR}})\).

\subsection{Low-Resolution Phase Predictor}
\vspace{-0.4em}

\paragraph{Network architecture.}
We employ an \emph{asymmetric} U-Net \(f_{\theta}\) whose encoder has four down-sampling blocks (stride~2 convolutions) and whose decoder has only two up-sampling blocks.  Consequently, the network output \(\hat{\phi}^{\text{LR}}=f_{\theta}(I)\) has spatial dimensions \(\tfrac14 H\times\tfrac14 W\).

\paragraph{Training loss.}
The predictor is trained with a mean-squared error (MSE) objective:
\begin{equation}
\mathcal{L}_{\text{LR}}(\theta)=
\mathbb{E}_{(I,\phi^{\text{HR}})}
\left[
\bigl\|
f_{\theta}(I)-\phi^{\text{LR}}
\bigr\|_2^{2}
\right].
\label{eq:lr_mse}
\end{equation}

\subsection{Diffusion Prior for High-Frequency Restoration}
\vspace{-0.4em}

\paragraph{Network architecture.}
The diffusion prior is a \emph{symmetric} U-Net \(\epsilon_{\omega}\) identical in depth on the encoder and decoder sides, operating at the full resolution \(H\times W\).

\paragraph{Forward diffusion process.}
Diffusion models first define a forward diffusion process to perturb the data distribution $p_{data}$ to a Gaussian distribution. Formally, the diffusion process follows an SDE $\mathrm{d}\vx_t=\vf(\vx_t)+g(t)\mathrm{d}\mathbf{w}$, where $\mathrm{d}\mathbf{w}$ is the Brownian motion and $t$ flows forward from $0$ to $T$. The calculated analytic solution of this diffusion process gives a transition distribution $p_t(\vx_t|\vx_0)=\cN(\vx_t|\alpha_t\vx_0,\sigma^2_t\mathbf{I})$, where $\alpha_t=e^{\int_0^t f(s)ds}$ and $\sigma_t^2=1-e^{-\int_0^t g(s)^2ds}$. In the typical variance-preserving diffusion schedule, $\vf$ and $g$ are designed such that $\lim_{t\to0}p_t(\vx)=p_{data}(\vx)$ and $\lim_{t\to T}p_t(\vx)=\mathcal{N}(\vx|\mathbf{0},\mI)$. 

\paragraph{Denoising score matching loss.}
Diffusion models sample data by reversing this diffusion process, where $\nabla_{\vx_t} \log p_t(\vx_t)$ is required. To learn this term, a neural network $s_\theta$ is trained to minimize an empirical risk by marginalizing $\nabla_{\vx_t} \log p_t(\vx_t|\vx_0)$, leading to the following loss:
\begin{equation*}
    \begin{aligned}                  
     L(\theta)=\E_{t\sim U(0,1), \epsilon\sim\cN(\mathbf{0},\mathbf{I})}\sum_{n=1}^{N}\|s_\theta(\alpha_t \vx_n+\sigma_t\epsilon, t)+\eps/{\sigma_t}\|^2. \\
    \end{aligned}
\end{equation*}
To further balance the diffusion loss at different $t$'s,  people usually adopt loss reweighing \cite{karras2022edm} or an alternate objective using $\epsilon$-prediction \cite{ho2020ddpm, nichol2021improved}, leading to the following well-known denoising score matching (DSM) loss:
\begin{equation*}
L(\theta,t)=\E_{\epsilon\sim\cN(\mathbf{0},\mathbf{I})}\sum_{n=1}^{N}\|s_\theta(\alpha_t \vx_n+\sigma_t\epsilon, t)-\eps\|^2.
\end{equation*}
where $s_\theta(\cdot,t)$ can be viewed as the learned score function at time $t$. 

\paragraph{Reverse SDE sampling.}
Given an estimated low-resolution phase map, we first corrupt it by the forward diffusion process to certain timestep $t$.
\( \tilde{\phi}_{T}=
\alpha_t\texttt{Up}\bigl(\hat{\phi}^{\text{LR}}\bigr)+\sigma_t\epsilon,\ 
\epsilon\!\sim\!\mathcal{N}(0,\mathbf{I}), \)
Starting from this noise-perturbed phase map, we sample from the diffusion model by applying a reverse-time SDE which reverses the diffusion process \cite{anderson1982reversesde}:
\begin{equation*}
    \mathrm{d}\vx_t=[\vf(\vx_t)-g(t)^2\nabla_{\vx_t} \log p_t(\vx_t)]\mathrm{d}t+\mathrm{d}\bar{\mathbf{w}},
\end{equation*}
where $\mathrm{d}\bar{\mathbf{w}}$ is the Brownian motion and $t$ flows forward from $T$ to $0$.

\section{Experiments}
\label{sec:experiments}
\begin{figure}[t]
  \centering
  \includegraphics[width=\linewidth]{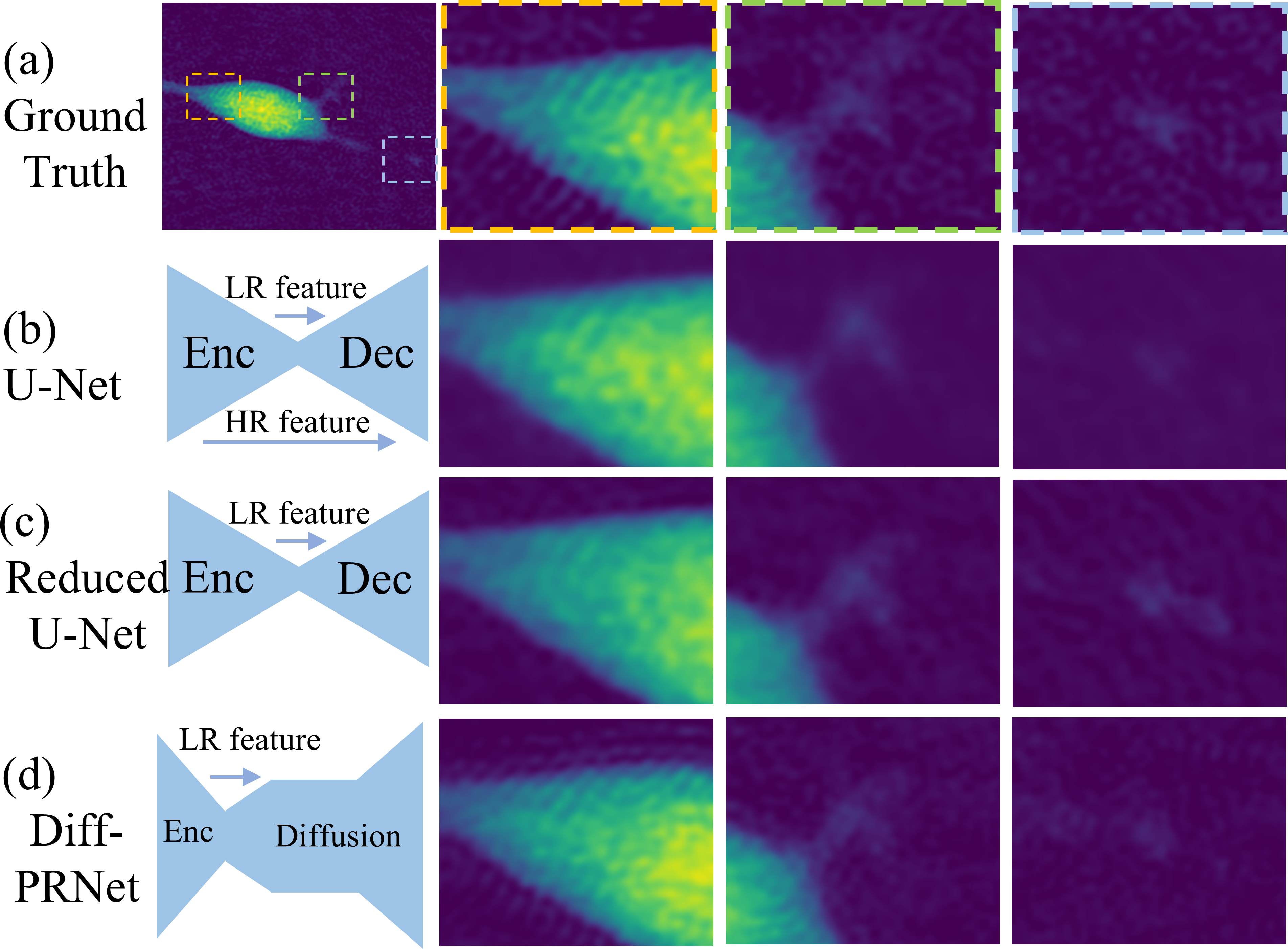}%
  \caption{\textbf{Qualitative comparison.}
    (a)~Ground-truth phase and zoom-in regions.
    (b)~U-Net results are overly smooth;
    (c)~Reduced U-Net restores better contrast but misses texture;
    (d)~DiffPR reconstructs sharp, noise-free details.}
  \label{fig:qualitative}
\end{figure}

\begin{table}[ht]
\centering
\renewcommand{\arraystretch}{1.2}
\caption{\bf Numerical Criteria for the Networks on Different Samples. The best results are in \textbf{bold}; $\times$ indicates that the model failed to converge and predicted a blank background.}
\begin{tabular}{ll S  S  S  S  S  S  S}
\toprule
 &  {Metric} & {B Cell} & {WBC} & {HeLa} & {3t3} \\
\midrule
\multirow{3}{*}{U-Net} 
& PSNR  & $\times$ & $\times$ & {28.59} & {34.41}\\ 
& SSIM  & $\times$ & $\times$ & {0.6119} & {0.7687}\\ 
& MAE  & $\times$ & $\times$ & {0.0317} & {0.0138}\\ 

\midrule
\multirow{3}{1cm}{Redu-ced U-Net} 
& PSNR  & {46.64} & {45.70} & {28.88} & {34.76}\\ 
& SSIM  & {0.9506} & {0.9145} & {0.7080} & {0.8278}\\ 
& MAE  & {0.0028} & {0.0035} & {0.0280} & {0.0119}\\ 

\midrule
\multirow{3}{1cm}{Diff-PRNet} 
& PSNR  & {\bfseries 47.59} & {\bfseries 46.81} & {\bfseries 29.48} & {\bfseries 36.96}\\ 
& SSIM  & {\bfseries 0.9659} & {\bfseries 0.9354} & {\bfseries 0.7393} & {\bfseries 0.8443}\\ 
& MAE  & {\bfseries 0.0025} & {\bfseries 0.0028} & {\bfseries 0.0249} & {\bfseries 0.0108}\\ 
\bottomrule

\end{tabular}
  \label{tab:main_exp}
\end{table} 

We evaluate \textbf{DiffPR} on four representative off-axis QPI datasets—\textit{B-Cell}, \textit{WBC}, \textit{HeLa}, and \textit{3T3}—and compare it against the canonical \textbf{U-Net} and our \textbf{Reduced U-Net}\footnote{U-Net without the deepest skip connections; see Fig.~\ref{fig:method}~(b).}.  
Results demonstrate that  
(1)~removing high-frequency skip connections already improves fidelity, and  
(2)~adding the diffusion prior further boosts both numerical and visual quality.

\subsection{Datasets and Metrics}

We construct four off-axis QPI datasets containing interferograms of
\textit{B Lymphocytes} (\num{385} samples),
\textit{white blood cells} (\num{312}),
\textit{HeLa} (\num{284}), and
\textit{3T3} cells (\num{298}).
Each interferogram is of size $1024\times1024$ pixels and is normalised to zero mean and unit variance.
Target phase maps are linearly rescaled to the $[0,1]$ range.

Following common practice, we randomly split each dataset into \SI{80}{\percent} training, \SI{10}{\percent} validation, and \SI{10}{\percent} test sets.
Performance is reported on the test split with
\textbf{PSNR} (\si{\decibel}),
\textbf{SSIM}, and
\textbf{MAE} (lower is better).

\subsection{Implementation Details}

\paragraph{Network variants.}
We train three models on all datasets:
(i)~the canonical \textbf{U-Net},
(ii)~the proposed \textbf{Reduced U-Net}, and
(iii)~our full \textbf{DiffPR} (which uses the Reduced U-Net as low-resolution predictor and a diffusion prior).

\paragraph{Training protocol for U-Net variants.}
Input--target pairs are randomly cropped to $512\times512$ to accelerate training and reduce GPU memory.
All models are optimised with Adam ($\beta_1{=}0.9,\;\beta_2{=}0.999$) at a learning rate of $1\times10^{-4}$ and a batch size of $16$.
For the vanilla U-Net we employ early stopping on the validation PSNR with a patience of $100$ epochs.

\paragraph{Diffusion prior.}
The diffusion model adopts a variance-preserving schedule~\cite{ho2020ddpm} with $T{=}1000$ timesteps and is trained for $800$k updates.
At inference we integrate the reverse SDE with a $100$-step predictor--corrector sampler (Alg.~\ref{alg:diffpr}).

\section{Quantitative Results}
Table \ref{tab:main_exp} presents PSNR, SSIM, and MAE across all four datasets. On \textit{B-Cell} and \textit{WBC} the baseline U-Net collapses to an all-zero prediction—marked by “$\times$” in the table—because both datasets contain large zero-phase backgrounds. The network minimises loss by staying in this local minimum, and standard gradient descent fails to escape. The experiments reveral serveral clear patterns:

\paragraph{ Removing high-frequency skips helps convergence and fidelity.}
Eliminating the deepest skip connection forces the decoder to rely on low-frequency structure rather than a spurious high-frequency shortcut. As a result, the \emph{Reduced U-Net} converges on every dataset and consistently outperforms the vanilla model. For example, on \textit{B-Cell} PSNR rises from failure to \SI{46.6}{dB}, and MAE decreases by \SI{13}{\percent}.

\paragraph{ A diffusion prior further closes the detail gap.}
Adding the diffusion stage boosts fidelity beyond the Reduced U-Net, especially on texture-rich or noisy samples. On \textit{WBC} we observe a +\SI{1.1}{dB} PSNR gain (from 45.7 dB to 46.8 dB) and a corresponding SSIM improvement (0.914 → 0.935). On \textit{B-Cell} MAE drops from 0.0028 to 0.0025. The iterative denoising process effectively reconstructs membrane granularity that deterministic decoders miss.

Overall, DiffPR delivers the highest PSNR on every dataset—up to +\SI{1.1}{dB} beyond the Reduced U-Net—and reduces MAE by as much as \SI{11}{\percent}. These results confirm that (i) suppressing direct high-frequency supervision avoids background-dominated minima, and (ii) coupling a low-frequency predictor with a diffusion prior is superior to conventional end-to-end optimisation for high-fidelity phase reconstruction.

\subsection{Qualitative Results}

Figure~\ref{fig:qualitative} illustrates visual differences on representative crops.  
Grey dashed boxes in the ground truth highlight three challenging regions: fine ridges, a low-contrast cytoplasmic boundary, and speckle-like noise.

\textbf{Vanilla U-Net (Fig.~\ref{fig:qualitative}b).}
Although edges are roughly aligned, high-frequency texture is missing; ridge amplitude is attenuated and intracellular speckles collapse into smooth blobs—classic symptoms of spectral bias.

\textbf{Reduced U-Net (Fig.~\ref{fig:qualitative}c).}
Removing top-level skips sharpens global shape and attenuates ringing, but residual blur remains around membrane folds and the faint cytoplasmic boundary is still over-smoothed.
This suggests that frequency decoupling helps preserve macroscopic structure yet cannot fully regenerate lost fine features.

\textbf{DiffPR (Fig.~\ref{fig:qualitative}d).}
Our method faithfully reconstructs the sinusoidal ridges, restores the weak boundary with correct phase gradient, and reproduces speckle patterning without hallucinated noise.
Because the diffusion prior progressively denoises from a low-frequency initialisation, it focuses its capacity on residual high-frequency modes rather than re-learning the full signal.
The result is a phase map visually indistinguishable from the ground truth while remaining free of artefacts.

These qualitative observations mirror the quantitative improvements in Table~\ref{tab:main_exp}, reinforcing that DiffPR achieves superior high-frequency fidelity without sacrificing low-frequency accuracy.

\section{Conclusion}
\label{sec:conclusion}

We have introduced \textbf{DiffPR}, a frequency‐decoupled framework for off‐axis quantitative phase imaging that couples a low‐resolution predictor with an unconditional diffusion prior.  
Our analysis revealed that the ubiquitous U\=/Net architecture, when endowed with deep skip connections, can overfit large background regions and stall in a low‐frequency local minimum.  
Removing the deepest skips (\emph{Reduced U\=/Net}) already alleviates this issue, confirming that direct supervision of high‐frequency features can be harmful.  
Building on this observation, DiffPR synthesises the missing high‐frequency residuals through a reverse SDE, explicitly exploiting the diffusion model’s bias toward detail recovery.

Extensive experiments on four cell datasets demonstrate that DiffPR surpasses both vanilla and Reduced U\=/Net baselines, improving PSNR by up to \SI{1.1}{dB} and lowering MAE by \SI{11}{\percent}.  
Qualitative comparisons further show that DiffPR reconstructs sub‐cellular ridges and speckle patterns with minimal artefacts.

\paragraph{Limitations and future work.}
DiffPR currently relies on a $100$‐step sampler, incurring longer inference time than a single forward pass.  
Future work will explore consistency distillation \cite{vanilladistillation, liu2023rectifiedflow, song2023consistencymodels, yin2024dmd, zheng2023fastno, liu2023instaflow, berthelot2023tract}, and higher‐order ODE solvers \cite{song2021ddim, lu2022dpmdeterminsitc, lu2022dpmplusplusdeterminsitc} to accelerate sampling.  
Moreover, extending the framework to volumetric phase tomography and to other ill‐posed inverse problems (e.g.\ fluorescence lifetime imaging) could broaden its impact.  
Finally, incorporating physics‐informed constraints into the diffusion prior is a promising direction for further improving accuracy and interpretability.

\bibliography{main}

\begin{thebibliography}{44}
\providecommand{\natexlab}[1]{#1}
\providecommand{\url}[1]{\texttt{#1}}
\expandafter\ifx\csname urlstyle\endcsname\relax
  \providecommand{\doi}[1]{doi: #1}\else
  \providecommand{\doi}{doi: \begingroup \urlstyle{rm}\Url}\fi

\bibitem[Anderson(1982)]{anderson1982reversesde}
Anderson, B.~D.
\newblock Reverse-time diffusion equation models.
\newblock \emph{Stochastic Processes and their Applications}, 12\penalty0 (3):\penalty0 313--326, 1982.

\bibitem[Berthelot et~al.(2023)Berthelot, Autef, Lin, Yap, Zhai, Hu, Zheng, Talbott, and Gu]{berthelot2023tract}
Berthelot, D., Autef, A., Lin, J., Yap, D.~A., Zhai, S., Hu, S., Zheng, D., Talbott, W., and Gu, E.
\newblock Tract: Denoising diffusion models with transitive closure time-distillation.
\newblock \emph{arXiv preprint arXiv:2303.04248}, 2023.

\bibitem[Bhaduri et~al.(2014)Bhaduri, Edwards, Pham, Zhou, Nguyen, Goddard, and Popescu]{bhaduri2014diffraction}
Bhaduri, B., Edwards, C., Pham, H., Zhou, R., Nguyen, T.~H., Goddard, L.~L., and Popescu, G.
\newblock Diffraction phase microscopy: principles and applications in materials and life sciences.
\newblock \emph{Advances in Optics and Photonics}, 6\penalty0 (1):\penalty0 57--119, 2014.

\bibitem[Chang et~al.(2020)Chang, Ryu, Jo, Choi, Min, and Park]{b3}
Chang, T., Ryu, D., Jo, Y., Choi, G., Min, H.-S., and Park, Y.
\newblock Calibration-free quantitative phase imaging using data-driven aberration modeling.
\newblock \emph{Optics Express}, 28\penalty0 (23):\penalty0 34835--34847, 2020.

\bibitem[Dhariwal \& Nichol(2021)Dhariwal and Nichol]{dhariwal2021diffusion}
Dhariwal, P. and Nichol, A.
\newblock Diffusion models beat gans on image synthesis.
\newblock \emph{Advances in Neural Information Processing Systems}, 2021.
\newblock arXiv:2105.05233.

\bibitem[Dieleman(2024)]{dieleman2024spectral}
Dieleman, S.
\newblock Diffusion is spectral autoregression.
\newblock Blog post, 2024.
\newblock URL \url{https://sander.ai/2024/09/02/spectral-autoregression.html}.
\newblock Accessed: 2025-06-12.

\bibitem[Ho et~al.(2020{\natexlab{a}})Ho, Jain, and Abbeel]{ho2020ddpm}
Ho, J., Jain, A., and Abbeel, P.
\newblock Denoising diffusion probabilistic models.
\newblock \emph{Advances in neural information processing systems}, 33:\penalty0 6840--6851, 2020{\natexlab{a}}.

\bibitem[Ho et~al.(2020{\natexlab{b}})Ho, Jain, and Abbeel]{ho2020denoising}
Ho, J., Jain, A., and Abbeel, P.
\newblock Denoising diffusion probabilistic models.
\newblock \emph{arXiv preprint arXiv:2006.11239}, 2020{\natexlab{b}}.

\bibitem[Karras et~al.(2022)Karras, Aittala, Aila, and Laine]{karras2022edm}
Karras, T., Aittala, M., Aila, T., and Laine, S.
\newblock Elucidating the design space of diffusion-based generative models.
\newblock \emph{Advances in neural information processing systems}, 35:\penalty0 26565--26577, 2022.

\bibitem[Kingma et~al.(2021)Kingma, Salimans, Poole, and Ho]{kingma2021vdm}
Kingma, D.~P., Salimans, T., Poole, B., and Ho, J.
\newblock Variational diffusion models.
\newblock In \emph{Advances in Neural Information Processing Systems (NeurIPS)}, 2021.
\newblock arXiv:2107.00630.

\bibitem[Lai et~al.(2021)Lai, Xiao, Xu, Fan, and Wei]{b4}
Lai, X., Xiao, S., Xu, C., Fan, S., and Wei, K.
\newblock Aberration-free digital holographic phase imaging using the derivative-based principal component analysis.
\newblock \emph{Journal of Biomedical Optics}, 26\penalty0 (4):\penalty0 046501, 2021.

\bibitem[Liu et~al.(2023{\natexlab{a}})Liu, Gong, and qiang liu]{liu2023rectifiedflow}
Liu, X., Gong, C., and qiang liu.
\newblock Flow straight and fast: Learning to generate and transfer data with rectified flow.
\newblock In \emph{The Eleventh International Conference on Learning Representations}, 2023{\natexlab{a}}.
\newblock URL \url{https://openreview.net/forum?id=XVjTT1nw5z}.

\bibitem[Liu et~al.(2023{\natexlab{b}})Liu, Zhang, Ma, Peng, et~al.]{liu2023instaflow}
Liu, X., Zhang, X., Ma, J., Peng, J., et~al.
\newblock Instaflow: One step is enough for high-quality diffusion-based text-to-image generation.
\newblock In \emph{The Twelfth International Conference on Learning Representations}, 2023{\natexlab{b}}.

\bibitem[Lu et~al.(2022{\natexlab{a}})Lu, Zhou, Bao, Chen, Li, and Zhu]{lu2022dpmdeterminsitc}
Lu, C., Zhou, Y., Bao, F., Chen, J., Li, C., and Zhu, J.
\newblock Dpm-solver: A fast ode solver for diffusion probabilistic model sampling in around 10 steps.
\newblock \emph{Advances in Neural Information Processing Systems}, 35:\penalty0 5775--5787, 2022{\natexlab{a}}.

\bibitem[Lu et~al.(2022{\natexlab{b}})Lu, Zhou, Bao, Chen, Li, and Zhu]{lu2022dpmplusplusdeterminsitc}
Lu, C., Zhou, Y., Bao, F., Chen, J., Li, C., and Zhu, J.
\newblock Dpm-solver++: Fast solver for guided sampling of diffusion probabilistic models.
\newblock \emph{arXiv preprint arXiv:2211.01095}, 2022{\natexlab{b}}.

\bibitem[Luhman \& Luhman(2021)Luhman and Luhman]{vanilladistillation}
Luhman, E. and Luhman, T.
\newblock Knowledge distillation in iterative generative models for improved sampling speed.
\newblock \emph{arXiv preprint arXiv:2101.02388}, 2021.

\bibitem[Nichol \& Dhariwal(2021)Nichol and Dhariwal]{nichol2021improved}
Nichol, A.~Q. and Dhariwal, P.
\newblock Improved denoising diffusion probabilistic models.
\newblock In \emph{International conference on machine learning}, pp.\  8162--8171. PMLR, 2021.

\bibitem[Niu et~al.(2020)Niu, Luo, Shu, Qu, Zhou, Ho, Zhao, and Zhou]{niu2020portable}
Niu, M., Luo, G., Shu, X., Qu, F., Zhou, S., Ho, Y.-P., Zhao, N., and Zhou, R.
\newblock Portable quantitative phase microscope for material metrology and biological imaging.
\newblock \emph{Photonics Research}, 8\penalty0 (7):\penalty0 1253--1259, 2020.

\bibitem[Pandey et~al.(2019)Pandey, Zhou, Bordett, Hunter, Glunde, Barman, Valdez, and Finck]{pandey2019integration}
Pandey, R., Zhou, R., Bordett, R., Hunter, C., Glunde, K., Barman, I., Valdez, T., and Finck, C.
\newblock Integration of diffraction phase microscopy and raman imaging for label-free morpho-molecular assessment of live cells.
\newblock \emph{Journal of biophotonics}, 12\penalty0 (4):\penalty0 e201800291, 2019.

\bibitem[Park et~al.(2006)Park, Popescu, Badizadegan, Dasari, and Feld]{Park:06}
Park, Y., Popescu, G., Badizadegan, K., Dasari, R.~R., and Feld, M.~S.
\newblock Diffraction phase and fluorescence microscopy.
\newblock \emph{Opt. Express}, 14\penalty0 (18):\penalty0 8263--8268, Sep 2006.
\newblock \doi{10.1364/OE.14.008263}.
\newblock URL \url{https://opg.optica.org/oe/abstract.cfm?URI=oe-14-18-8263}.

\bibitem[Park et~al.(2018)Park, Depeursinge, and Popescu]{park2018quantitative}
Park, Y., Depeursinge, C., and Popescu, G.
\newblock Quantitative phase imaging in biomedicine.
\newblock \emph{Nature photonics}, 12\penalty0 (10):\penalty0 578--589, 2018.

\bibitem[Popescu et~al.(2006)Popescu, Ikeda, Dasari, and Feld]{popescu2006diffraction}
Popescu, G., Ikeda, T., Dasari, R.~R., and Feld, M.~S.
\newblock Diffraction phase microscopy for quantifying cell structure and dynamics.
\newblock \emph{Optics letters}, 31\penalty0 (6):\penalty0 775--777, 2006.

\bibitem[Rahaman et~al.(2019)Rahaman, Baratin, Arpit, Draxler, Lin, Hamprecht, Bengio, and Courville]{pmlr-v97-rahaman19a}
Rahaman, N., Baratin, A., Arpit, D., Draxler, F., Lin, M., Hamprecht, F., Bengio, Y., and Courville, A.
\newblock On the spectral bias of neural networks.
\newblock In \emph{Proceedings of the 36th International Conference on Machine Learning}, Proceedings of Machine Learning Research, pp.\  5301--5310. PMLR, 2019.
\newblock URL \url{https://proceedings.mlr.press/v97/rahaman19a.html}.

\bibitem[Rivenson et~al.(2018)Rivenson, Zhang, G{\"u}nayd{\i}n, Teng, and Ozcan]{rivenson2018phase}
Rivenson, Y., Zhang, Y., G{\"u}nayd{\i}n, H., Teng, D., and Ozcan, A.
\newblock Phase recovery and holographic image reconstruction using deep learning in neural networks.
\newblock \emph{Light: Science \& Applications}, 7\penalty0 (2):\penalty0 17141--17141, 2018.

\bibitem[Rombach et~al.(2022)Rombach, Blattmann, Lorenz, Esser, and Ommer]{rombach2022latent}
Rombach, R., Blattmann, A., Lorenz, D., Esser, P., and Ommer, B.
\newblock High-resolution image synthesis with latent diffusion models.
\newblock In \emph{IEEE/CVF Conference on Computer Vision and Pattern Recognition (CVPR)}, 2022.
\newblock arXiv:2112.10752.

\bibitem[Ronneberger et~al.(2015)Ronneberger, Fischer, and Brox]{ronneberger2015u}
Ronneberger, O., Fischer, P., and Brox, T.
\newblock U-net: Convolutional networks for biomedical image segmentation.
\newblock In \emph{International Conference on Medical image computing and computer-assisted intervention}, pp.\  234--241. Springer, 2015.

\bibitem[Russo(2025)]{gianluca2025spectral}
Russo, G.
\newblock Diffusion is autoregression in the frequency domain.
\newblock Blog post, 2025.
\newblock URL \url{https://gianluca.ai/diffusion-is-frequency-autoregression/}.
\newblock Accessed: 2025-06-12.

\bibitem[Ryu et~al.(2021)Ryu, Kim, Lim, Min, Yoo, Cho, and Park]{ryu2021label}
Ryu, D., Kim, J., Lim, D., Min, H.-S., Yoo, I.~Y., Cho, D., and Park, Y.
\newblock Label-free white blood cell classification using refractive index tomography and deep learning.
\newblock \emph{BME Frontiers}, 2021, 2021.

\bibitem[Saharia et~al.(2022{\natexlab{a}})Saharia, Chan, Chang, Lee, Ho, Salimans, Fleet, and Norouzi]{saharia2022palette}
Saharia, C., Chan, W., Chang, H., Lee, C., Ho, J., Salimans, T., Fleet, D.~J., and Norouzi, M.
\newblock Palette: Image-to-image diffusion models.
\newblock In \emph{ACM SIGGRAPH Conference Proceedings}, 2022{\natexlab{a}}.
\newblock arXiv:2111.05826.

\bibitem[Saharia et~al.(2022{\natexlab{b}})Saharia, Ho, Chan, Salimans, Fleet, and Norouzi]{saharia2021sr3}
Saharia, C., Ho, J., Chan, W., Salimans, T., Fleet, D.~J., and Norouzi, M.
\newblock Image super-resolution via iterative refinement.
\newblock In \emph{IEEE/CVF Conference on Computer Vision and Pattern Recognition (CVPR)}, 2022{\natexlab{b}}.
\newblock arXiv:2104.07636.

\bibitem[Shi et~al.(2019)Shi, Zhu, Wang, Song, and Guo]{b5}
Shi, J., Zhu, X., Wang, H., Song, L., and Guo, Q.
\newblock Label enhanced and patch based deep learning for phase retrieval from single frame fringe pattern in fringe projection 3d measurement.
\newblock \emph{Optics express}, 27\penalty0 (20):\penalty0 28929--28943, 2019.

\bibitem[Song et~al.(2021)Song, Meng, and Ermon]{song2021ddim}
Song, J., Meng, C., and Ermon, S.
\newblock Denoising diffusion implicit models.
\newblock In \emph{International Conference on Learning Representations}, 2021.
\newblock URL \url{https://openreview.net/forum?id=St1giarCHLP}.

\bibitem[Song et~al.(2020)Song, Sohl-Dickstein, Kingma, Kumar, Ermon, and Poole]{song2020score}
Song, Y., Sohl-Dickstein, J., Kingma, D.~P., Kumar, A., Ermon, S., and Poole, B.
\newblock Score-based generative modeling through stochastic differential equations.
\newblock In \emph{International Conference on Learning Representations}, 2020.

\bibitem[Song et~al.(2023)Song, Dhariwal, Chen, and Sutskever]{song2023consistencymodels}
Song, Y., Dhariwal, P., Chen, M., and Sutskever, I.
\newblock Consistency models, 2023.
\newblock URL \url{https://arxiv.org/abs/2303.01469}.

\bibitem[Tancik et~al.(2020)Tancik, Srinivasan, Mildenhall, Fridovich-Keil, Raghavan, Singhal, Ramamoorthi, Barron, and Ng]{tancik2020fourier}
Tancik, M., Srinivasan, P.~P., Mildenhall, B., Fridovich-Keil, S., Raghavan, N., Singhal, U., Ramamoorthi, R., Barron, J.~T., and Ng, R.
\newblock Fourier features let networks learn high frequency functions in low dimensional domains.
\newblock In \emph{Advances in Neural Information Processing Systems}, 2020.
\newblock URL \url{https://arxiv.org/abs/2006.10739}.

\bibitem[Wang et~al.(2018)Wang, Lyu, and Situ]{Wang:18}
Wang, H., Lyu, M., and Situ, G.
\newblock eholonet: a learning-based end-to-end approach for in-line digital holographic reconstruction.
\newblock \emph{Opt. Express}, 26\penalty0 (18):\penalty0 22603--22614, Sep 2018.
\newblock \doi{10.1364/OE.26.022603}.
\newblock URL \url{https://opg.optica.org/oe/abstract.cfm?URI=oe-26-18-22603}.

\bibitem[Wang et~al.(2019{\natexlab{a}})Wang, Li, Kemao, Di, and Zhao]{Wang:19}
Wang, K., Li, Y., Kemao, Q., Di, J., and Zhao, J.
\newblock One-step robust deep learning phase unwrapping.
\newblock \emph{Opt. Express}, 27\penalty0 (10):\penalty0 15100--15115, May 2019{\natexlab{a}}.
\newblock \doi{10.1364/OE.27.015100}.
\newblock URL \url{https://opg.optica.org/oe/abstract.cfm?URI=oe-27-10-15100}.

\bibitem[Wang et~al.(2019{\natexlab{b}})Wang, Li, Kemao, Di, and Zhao]{b2}
Wang, K., Li, Y., Kemao, Q., Di, J., and Zhao, J.
\newblock One-step robust deep learning phase unwrapping.
\newblock \emph{Optics express}, 27\penalty0 (10):\penalty0 15100--15115, 2019{\natexlab{b}}.

\bibitem[Xu et~al.(2019)Xu, Zhang, Luo, Xiao, and Ma]{xu2019frequency}
Xu, Z.-Q.~J., Zhang, Y., Luo, T., Xiao, Y., and Ma, Z.
\newblock Frequency principle: Fourier analysis sheds light on deep neural networks.
\newblock \emph{arXiv preprint arXiv:1901.06523}, 2019.
\newblock URL \url{https://arxiv.org/abs/1901.06523}.

\bibitem[Yao et~al.(2020)Yao, Shu, and Zhou]{b6}
Yao, Y., Shu, X., and Zhou, R.
\newblock Deep learning based phase retrieval in quantitative phase microscopy.
\newblock In \emph{Unconventional Optical Imaging II}, volume 11351, pp.\  76--80. SPIE, 2020.

\bibitem[Yin et~al.(2024)Yin, Gharbi, Zhang, Shechtman, Durand, Freeman, and Park]{yin2024dmd}
Yin, T., Gharbi, M., Zhang, R., Shechtman, E., Durand, F., Freeman, W.~T., and Park, T.
\newblock One-step diffusion with distribution matching distillation.
\newblock In \emph{Proceedings of the IEEE/CVF conference on computer vision and pattern recognition}, pp.\  6613--6623, 2024.

\bibitem[Zhang et~al.(2018)Zhang, Guan, Shen, Wang, Hu, Wang, He, and Xie]{Zhang:18}
Zhang, G., Guan, T., Shen, Z., Wang, X., Hu, T., Wang, D., He, Y., and Xie, N.
\newblock Fast phase retrieval in off-axis digital holographic microscopy through deep learning.
\newblock \emph{Opt. Express}, 26\penalty0 (15):\penalty0 19388--19405, Jul 2018.
\newblock \doi{10.1364/OE.26.019388}.
\newblock URL \url{https://opg.optica.org/oe/abstract.cfm?URI=oe-26-15-19388}.

\bibitem[Zhang et~al.(2021)Zhang, Noack, Vagovic, Fezzaa, Garcia-Moreno, Ritschel, and Villanueva-Perez]{zhang2021phasegan}
Zhang, Y., Noack, M.~A., Vagovic, P., Fezzaa, K., Garcia-Moreno, F., Ritschel, T., and Villanueva-Perez, P.
\newblock Phasegan: a deep-learning phase-retrieval approach for unpaired datasets.
\newblock \emph{Optics express}, 29\penalty0 (13):\penalty0 19593--19604, 2021.

\bibitem[Zheng et~al.(2023)Zheng, Nie, Vahdat, Azizzadenesheli, and Anandkumar]{zheng2023fastno}
Zheng, H., Nie, W., Vahdat, A., Azizzadenesheli, K., and Anandkumar, A.
\newblock Fast sampling of diffusion models via operator learning.
\newblock In \emph{International conference on machine learning}, pp.\  42390--42402. PMLR, 2023.

\end{thebibliography}
\bibliographystyle{icml2025}

\end{document}